%% file: Tsitseika-FF.tex
\documentclass[11pt,twoside]{gsp}
\input somedef

\usepackage{amssymb,amsmath,graphics,epsfig,eepic}
\usepackage{wrapfig}
\usepackage{color}
\def\i{\mathrm{i\,}}

\def\diag{\text{diag\,}}

\def\openone{\leavevmode\hbox{\small1\kern-3.3pt\normalsize1}}

\def\bbbz{\Bbb Z}
\def\bbbd{\Bbb D}

\begin{document}

\thispagestyle{plain}

\title{ON THE SOLITON SOLUTIONS OF A FAMILY OF TZITZEICA  EQUATIONS}
\author{Corina N. Babalic$^{1,2}$, Radu Constantinescu$^{2}$ and  Vladimir S. Gerdjikov$^{3}$}

\date{}

\maketitle

\vspace*{-5mm}
\comm{Communicated by Metin G\"urses}\

\begin{abstract}
We analyze several types of soliton solutions to a family of Tzitzeica equations. To this end we  use two methods for deriving the soliton solutions:
the dressing method and Hirota method. The dressing method allows us to derive two types of soliton solutions. The first type corresponds to a set of 6
symmetrically situated discrete eigenvalues of the Lax operator $L$; to each soliton of the second type one relates a set of 12 discrete eigenvalues of $L$.
We also outline how one can construct general $N$ soliton solution containing $N_1$ solitons of first type and $N_2$ solitons of second type, $N=N_1+N_2$.
The possible singularities of the solitons and the effects of change of variables that relate the different members of Tzitzeica family  equations are briefly discussed.
All equations allow quasi-regular as well as singular soliton solutions.



\end{abstract}
\vspace*{-10mm}
\tableofcontents

\label{first}

\section{Introduction }

In the present paper we continue our investigations of the famous equation due to the Romanian mathematician Gheorghe Tzitzeica
\footnote{The name of the famous Romanian mathematician contains the Romanian letter \c T, which may be spelled as Tz. The factor 2 in eq. (\ref{eq:Ts12})
can be easily removed, but is kept for historical reasons.  },
which we call now as Tzitzeica 1 equation \cite{Tzi1,Tzi2} and a closely related equation which we call Tzitzeica 2; in what follows we
will denote them by \c T1 and \c T2.
It was initially proposed as an equation describing special surfaces in differential geometry for which the ratio $K/d^4$ is constant,
where $K$ is the Gauss curvature of the surface and $d$ is the distance from the origin to the tangent plane at the given point.
Later on it turned out that the equation has wider importance, being nowadays used as an important evolutionary equation in nonlinear
dynamics.

The explicit form of \c T1 and \c T2 equations is
\begin{equation}\label{eq:Ts12}\begin{split}
2 \frac{\partial^2 \phi_1 }{\partial \xi  \partial \eta } &= \e^{2\phi_1} - \e^{-4\phi_1} \qquad 
2 \frac{\partial^2 \phi_2 }{\partial \xi  \partial \eta } = - (\e^{2\phi_2} - \e^{-4\phi_2} )
\end{split}\end{equation}
i.e. \c T1 and \c T2 have different  signs in the right hand sides. The transition between T1 and T2 can be  performed by several simple
changes of variables (see below), some of which  substantially modify the singularity properties of their solutions.

Tzitzeica equations attracted a lot of attention at the end of the '70-ies when for some time it was believed,
that it is the only known equation, allowing a finite number of higher integrals of motion \cite{DoBu}. Soon however,
it was proved that in fact, it possesses, like the other soliton equations,  an infinite number of integrals of motion \cite{ZhSh}.
Next it was discovered that the equation has a hidden $\mathbb{Z}_3$ symmetry, which becomes evident in its Lax representation
\cite{Mik1,mik}. This important discovery led Mikhailov to the notion of the reduction group and to the family of two-dimensional
Toda field theories (TFT) related to the $sl(n)$ algebras \cite{mik}. Soon after it was established that: i) 2-dimensional TFT can be
related to any of the simple Lie algebras \cite{OlPerMi,Olive,DriSok*84,LezSav};
ii) other classes of integrable NLEE may also possess such symmetries \cite{DriSok*84,VSG-88,VG2,CIL*10};
and iii) the expansions over the squared solutions and the theory of their recursion operators can be constructed \cite{GuKV,Yan2012,GeYa1}.

In  previous papers \cite{BJGA,PAUC}  we presented in the derivation of  the soliton solutions  of \c T1.
Both versions of  Tzitzeica equation allow Lax representation proposed by Mikhailov  \cite{Mik1,mik}.
This allows one to apply the dressing method of Zakharov-Shabat-Mikhailov \cite{ZaSha,mik,ZaMi} for calculating their soliton solutions.
In fact all these equations are particular examples of  2-dimensional Toda field theories (TFT) \cite{mik,DriSok*84,OlPerMi,Olive,LezSav}.
They all can be solved exactly using the inverse scattering method \cite{FadTakh87,ZMNP,GYaV*08}.

In the present paper we start with the analysis of a more general class of equations, which we call Tzitzeica family equations. Their
general form is
\begin{equation}\label{eq:TzF}\begin{split}
2 \frac{\partial^2 \phi }{\partial \xi  \partial \eta } =  \epsilon_1 c_1^2 \e^{2\phi} + \epsilon_2 c_2^2 \e^{-4\phi}
\end{split}\end{equation}
where $\epsilon_1^2 = \epsilon_2^2 =1$ and $c_1$ and $c_2$ are some positive real constants. Obviously equation \c T1 (resp. equation  \c T2)
is obtained from (\ref{eq:TzF}) by putting $\epsilon_1 =1$, $\epsilon_2=-1$, $c_1=c_4=1$ (resp. $\epsilon_1 =-1$, $\epsilon_2=1$, $c_1=c_4=1$).
We will call \c T3 and \c T4 the equations
\begin{equation}\label{eq:Ts34}\begin{split}
2 \frac{\partial^2 \phi_3 }{\partial \xi  \partial \eta } &= -\e^{2\phi_3} - \e^{-4\phi_3} \qquad
2 \frac{\partial^2 \phi_4 }{\partial \xi  \partial \eta } = \e^{2\phi_4} + \e^{-4\phi_4} .
\end{split}\end{equation}
which follow from (\ref{eq:TzF}) with $\epsilon_1 =\epsilon_2=-1$, $c_1=c_4=1$ and $\epsilon_1 =\epsilon_2=1$, $c_1=c_4=1$ respectively.

The paper is organized as follows. In Section 2  we  study a class of changes of variables that interrelate different members of Tzitzeica family.
We shall see that \c T1 -- \c T4 equations allow Lax representations so they can be solved exactly by the inverse scattering
method, \cite{Mik1,noi}.  In Section 3 the Zakharov-Shabat dressing method \cite{ZaSha}, adapted to systems with deep reductions,
\cite{mik,Mik1}  is used to construct their soliton solutions.
As a result we derive the soliton solutions of first and second types and analyze their
singularities.  Indeed, we find that even the simplest one-soliton solutions of first type  may have an infinite number of  singularities for finite
values of $\xi, \eta$. Such singularities are characteristic also for other soliton-type equations, e.g. for Liouville equation \cite{APPol1,APPol2,Pogr,Pogr2},
for sinh-Gordon equation and others, see e.g. \cite{Pogr,Fan,Mats} and the references therein. At the same time,
using an appropriate change of variables we obtain a solution having singularities at only two points which we call `quasi-regular'.
In Section 4 we  outline how the dressing formalism can be extended to derive the $N$-soliton solution of the considered
model with $N_1$ solitons of first type and $N_2$ solitons of second type, $N=N_1+N_2$.
 In Section 5 we demonstrate how Hirota method can be applied for deriving the soliton solutions of Tzitzeica eqs. and
show that it  results  compatible with the ones of the dressing method.  In Section 6 we briefly outline the spectral properties of
the Lax operators $L$. We demonstrate that the resolvent of $L$ has pole singularities that coincide with the poles of the dressing factor and its inverse.
We end by a discussion and conclusions.

%\section{Tzitzeica I versus Tzitzeica II equation}
\section{Lorentz (Anti-)Invariance in 2-Dimensions}

Obviously each of the TFT mentioned above can be viewed as a member of a hierarchy of  NLEE which can be solved
by applying the ISM to the corresponding Lax operator. However the Lorentz invariance  singles out the TFT models
from  all the other members of NLEE in the hierarchy. Indeed,  the TFT models allow changes of variables which may
drastically change, as we shall demonstrate below, the properties of the soliton solutions.

\subsection{Changes of Variables and the Lorentz (Anti-)Invariance }

Let us now consider how simple linear change of variables
\begin{equation}\label{eq:A0}\begin{aligned}
\vec{Y}' = A \vec{Y}, \qquad \vec{Y}' = \left(\begin{array}{c} \xi' \\ \eta'  \end{array}\right) ,
\qquad \vec{Y} = \left(\begin{array}{c} \xi \\ \eta  \end{array}\right) , \qquad A = \left(\begin{array}{cc} a & b  \\ c & d  \end{array}\right)
\end{aligned}\end{equation}
affect the solutions of Tzitzeica eqs. Obviously this transformations have to preserve, up to a sign,
$\frac{\partial^2 }{ \partial \xi \partial \eta }$ which means that
\begin{equation}\label{eq:A1}\begin{split}
A^T \sigma_1 A =  \pm \sigma_1, \qquad \sigma_1=\left(\begin{array}{cc} 0 & 1 \\ 1 & 0  \end{array}\right)
\end{split}\end{equation}
which is equivalent to the relations:
\begin{equation}\label{eq:A2}\begin{split}
ac = bd = 0, \qquad ad+bc = \pm 1.
\end{split}\end{equation}
These relations are satisfied in two cases
\begin{equation}\label{eq:A3}\begin{split}
\mbox{1)} \quad A_1^\pm = \left(\begin{array}{cc}a & 0 \\ 0 & \pm 1/a \end{array}\right), \qquad
\mbox{2)} \quad A_2^\pm = \left(\begin{array}{cc}0 & b \\ \pm 1/b & 0 \end{array}\right).
\end{split}\end{equation}

Here $a$ and $b$ can be, in general, arbitrary complex numbers. However, below we will consider two cases:
i) $a$ and $b$ -- real and ii) $a$ and $b$ -- purely imaginary.

Second class of transformations involves shifts of the field $\phi $
\begin{equation}\label{eq:phi'}\begin{split}
\phi (\xi,\eta) \to \phi' (\xi,\eta) = \phi  (\xi,\eta) - \ln c_0 + s_0 \frac{\pi \i}{2}
\end{split}\end{equation}
where $c_0>0$ is a real constant and $s_0$ takes the values $0$ and $1$. If $s_0=0$  and $c_0=c_1$ then \c T1 goes into
\begin{equation}\label{eq:Ts1'}\begin{split}
2 \frac{\partial^2 \phi'_1 }{ \partial \xi \partial \eta } = c_1^2 \e^{2\phi_1'} -  c_1^{-4} \e^{-4\phi_1'}
\end{split}\end{equation}
and similar expression for the \c T2 equation for $\phi_2$, but with opposite signs for the terms in the right hand side.

If we now choose $s_0=1$  and $c_0=c_1$ then \c T1 goes into
\begin{equation}\label{eq:Ts3'}\begin{split}
2 \frac{\partial^2 \phi'_1 }{ \partial \xi \partial \eta } = - c_1^2 \e^{2\phi_1'} -  c_1^{-4} \e^{-4\phi_1'}
\end{split}\end{equation}
which for $c_1=1$ coincides with \c T3 equation. We have listed  the results of several such transformations in Table 1.

\begin{table}
  \centering
\begin{tabular}{|l|c|c|c|c|}
  \hline
  % after \\: \hline or \cline{col1-col2} \cline{col3-col4} ...
    &  T1  &  T2 &   T3 &  T4  \\ \hline
$ A_{1,2}^+ $, $s_0=0$  &  T1 &  T2 &  T3 &  T4 \\
$ A_{1,2}^- $, $s_0=0$ &  T2 &  T1 &  T4 &  T3 \\
$ A_{1,2}^+ $, $s_0=1$ &  T3 &  T4 &  T1 &  T2 \\
$ A_{1,2}^- $, $s_0=1$ &  T4 &  T3 &  T2 &  T1 \\
  \hline
\end{tabular}
  \caption{Changes of variables that relate different members of Tzitzeica family equations.}\label{tab:1}
\end{table}

\subsection{The Lax Representation of \c T2 Equation}
Since different members of Tzitzeica family are related by changes of variables (see Table \ref{tab:1}), then it will be enough to
consider the Lax representation and soliton solutions of only one of them, say the second equation in (\ref{eq:Ts12}) \c T2.
 It admits the following Lax representation
\begin{equation}\label{Lax dressed}\begin{split}
L_1 \Psi (\xi,\eta, \lambda )&\equiv \i \frac{\partial\Psi(\xi,\eta, \lambda )}{ \partial \xi } + 2 \i \phi_\xi H_0 \Psi (\xi,\eta, \lambda )+\lambda \mathcal{J} \Psi (\xi,\eta, \lambda )=0 \\
L_2 \Psi(\xi,\eta, \lambda ) & \equiv \i \frac{\partial \Psi(\xi,\eta, \lambda )}{ \partial \eta } + \lambda^{-1} V_1 \Psi (\xi,\eta, \lambda )=0
\end{split}\end{equation}
where
\begin{equation}\label{eq:L2t}\begin{split}
H_0 =\left(\begin{array}{ccc} 1 & 0 & 0 \\ 0 & 0  & 0\\ 0 & 0 & -1  \end{array}\right) , \quad \mathcal{J} =\left(\begin{array}{ccc} 0 & 1 & 0 \\ 0 & 0  & 1\\ 1 & 0 & 0  \end{array}\right)  \quad
V_1(\xi,\eta)= \left(\begin{array}{ccc} 0 & 0 & \e^{-4\phi} \\  \e^{2\phi}  & 0 & 0 \\ 0 & \e^{2\phi} &  0  \end{array}\right).
\end{split}\end{equation}

The reductions of the Lax pair for  \c T2 equation  are similar but not the same as for the well known \c T1 eq. \cite{BJGA}:
\begin{enumerate}
  \item $\bbbz_3$-reduction
  \begin{equation}\label{eq:z3}\begin{split}
   Q^{-1} \Psi (\xi,\eta, \lambda) Q = \Psi (\xi,\eta, q\lambda) , \quad Q= \left(\begin{array}{ccc} 1 & 0 & 0 \\
   0 & q & 0 \\ 0 & 0 & q^2  \end{array}\right), \quad q=e^{2\pi \i/3}
  \end{split}\end{equation}
  which restricts $H_0$, $\mathcal{J}$ and $V_1$ by
  \begin{equation}\label{eq:z3-uv}
  Q^{-1} H_0 Q = H_0, \qquad Q^{-1} \mathcal{J} Q = q \mathcal{J}, \qquad Q^{-1} V_1 Q = q^{-1}V_1.
\end{equation}
  These conditions are satisfied identically.

  \item First $\bbbz_2$-reduction
  \begin{equation}\label{eq:z2a}\begin{split}
   \Psi ^*(\xi,\eta, -\lambda^* ) = \Psi (\xi,\eta, \lambda )
  \end{split}\end{equation}
  i.e.
  \begin{equation}\label{eq:z2a-uv}\begin{split}
  H_0 = H_0^*, \qquad J_1 = J_1^*,  \qquad   V_1 = V_1^*
  \end{split}\end{equation}
which means that $\phi=\phi^*$.

  \item Second $\bbbz_2$-reduction
  \begin{equation}\label{eq:z2b}\begin{split}
   A_0^{-1} \Psi^\dag (\xi, \eta, \lambda^* )A_0 = \Psi^{-1} (\xi,\eta, \lambda ), \qquad A_0 = \left(\begin{array}{ccc}
   0 & 0 & 1 \\ 0 & 1 & 0 \\ 1 & 0 & 0 \end{array}\right)
  \end{split}\end{equation}
  i.e.
  \begin{equation}\label{eq:z2b-uv}\begin{split}
  A_0^{-1} H_0^\dag A_0 &= -H_0, \qquad A_0^{-1} \mathcal{J}^\dag A_0 = \mathcal{J}, \qquad A_0^{-1} V_1^\dag A_0 = V_1.
  \end{split}\end{equation}

\end{enumerate}

\section{The Dressing Method and Dressing Factors for \c T2 Equation}

Let us start with a Lax representation of the form
\begin{equation}\label{Lax naked}\begin{split}
L_{10} \Psi_0 &\equiv \i \frac{\partial\Psi_0}{ \partial \xi } +\lambda \mathcal{J} \Psi_0 =0 \qquad
L_{20} \Psi_0  \equiv \i \frac{\partial \Psi_0}{ \partial \eta } + \lambda^{-1} \mathcal{J}^2 \Psi_0=0.
\end{split}\end{equation}

The fundamental solution $\Psi_0(\xi,\eta,\lambda)$,
known also as  the `naked' solution,  has as potential the trivial solution of \c T2 equation:
$\phi_0(\xi,\eta)=0$

The 'dressed' Lax pair, given by (\ref{Lax dressed}), admits the "dressed" fundamental solution  $\Psi(\xi,\eta,\lambda)$, with the potential the nontrivial solution $\phi(\xi,\eta)$.

The fundamental solutions  $\Psi$ and $\Psi_0$ are related by the dressing factor  $u(\xi,\eta,\lambda)$,
\begin{equation}\label{eq:u01}\begin{split}
\Psi(\xi,\eta,\lambda) = u(\xi,\eta,\lambda) \Psi_0(\xi,\eta,\lambda)u^{-1}_+(\lambda), \quad u_+(\lambda)=\lim_{\xi \rightarrow \infty}  u(\xi,\eta,\lambda)
\end{split}\end{equation}
which means that $u(\xi,\eta,\lambda) $ must satisfy
\begin{equation}\label{eq pt u}\begin{split}
\i\frac{\partial u} { \partial \xi }  + 2 \i \phi_\xi H_0 u(\xi,\eta,\lambda)  +\lambda \left[  \mathcal{J} , u(\xi,\eta,\lambda) \right]=0 \\
\i \frac{\partial u} { \partial \eta }   +\frac{1}{\lambda} V_1 u(\xi,\eta,\lambda) - \frac{1}{\lambda} u(\xi,\eta,\lambda)  \mathcal{J}^2
=0.
\end{split}\end{equation}

Since both Lax pairs (the dressed (\ref{Lax dressed}) and the naked one (\ref{Lax naked})) satisfy the three reductions, then also the dressing factor must satisfy them
\begin{equation}\label{eq:u-reds}\begin{aligned}
& \mbox{a)} & \; Q^{-1} u(\xi,\eta,\lambda ) Q &= u(\xi,\eta, q\lambda ), &\; & \mbox{b)}
& \; u^*(\xi,\eta,-\lambda^* ) = u(\xi,\eta,\lambda )  \\
& \mbox{c)} & \; A_0^{-1} u^\dag (\xi,\eta,\lambda^* ) A_0 &= u^{-1} (\xi,\eta,\lambda )
\end{aligned}\end{equation}
where  $A_0$ is defined by eq. (\ref{eq:z2b}).

\subsection{One Soliton Solution of First Type  }

 A natural  anzatz for the dressing factor with simple poles in $\lambda$ is \cite{mik}
\begin{equation}\label{eq:u0}\begin{split}
u(\xi,\eta,\lambda) =
\openone + \frac{1}{3}\left( \frac{A_1 }{\lambda -\lambda_1} + \frac{Q^{-1}A_1Q }{\lambda q^2 -\lambda_1}  +
\frac{Q^{-2} A_1Q^2 }{\lambda q -\lambda_1}\right)
\end{split}\end{equation}
where $A_1(\xi,\eta)$ is a $3 \times 3$ degenerate matrix of the form
\begin{equation}\label{eq:AA1}\begin{split}
A_1(\xi,\eta) = |n(\xi,\eta)\rangle \langle m^T(\xi,\eta)| \qquad 
(A_1)_{ij}(\xi,\eta) = n_i(\xi,\eta)m_j(\xi,\eta) .
\end{split}\end{equation}

The first reduction (\ref{eq:u-reds}a) on $u(x,t,\lambda)$  is automatically satisfied by the anzatz (\ref{eq:u0}).
The second reduction (\ref{eq:u-reds}b) leads to
\begin{equation}\label{eq:u1}\begin{split}
\frac{ \eta_{j-k} n_k m_j}{\lambda^3 - \lambda_1^{*,3}} = -\frac{ \rho_{j-k} n_k^* m_j^*}{\lambda^3 + \lambda_1^{*,3}} .
\end{split}\end{equation}
Here and below $j-k$ is understood modulo $3$ and
\begin{equation}\label{eq:A11}\begin{aligned}
\eta_0 & = \lambda_1^{2}, &\qquad \eta_1 & = \lambda \lambda_1, &\qquad \eta_2 & = \lambda^2\\
\rho_0 & = \lambda_1^{*,2}, &\qquad \rho_1 & = -\lambda \lambda_1^*, &\qquad \rho_2 & = \lambda^2.
\end{aligned}\end{equation}
In addition we must have
\[ \lambda_1^{*,3} =-\lambda_1^3\]
and
\begin{equation}\label{eq:A12}\begin{split}
\lambda_1^{*,2} n_i^* m_i^* = -\lambda_1^2 n_im_i, \quad \lambda_1^{*} n_i^* m_{i+1}^* = \lambda_1 n_i m_{i+1},\quad
 n_i^* m_{i+2}^* =  -n_im_{i+2}
\end{split}\end{equation}
where again all matrix indices are understood modulo 3. These relations can be rewritten as
\begin{equation}\label{eq:in1}\begin{split}
\arg n_i+ \arg m_i &=\frac{\pi}{2}- 2 \arg \lambda_1, \qquad \arg n_i + \arg m_{i+1} =- \arg \lambda_1 \\
\arg n_i + \arg m_{i+2} &=\frac{\pi}{2}, \qquad  \arg \lambda_1 = (2 k+1)\frac{\pi }{6}, \quad k=0,1, ... ,5.
\end{split}\end{equation}
So we can consider with no limitations that $\lambda_1=-\lambda_1^*$ and $A_1= -A_1^*$.
More specifically we will assume that the vector $\langle m^T(\xi, \eta)|$ is real, while  the vector $|n(\xi, \eta)\rangle$ has purely imaginary components.

The third reduction (\ref{eq:u-reds}c) on $u(x,t,\lambda)$  can be put in the form
\begin{equation}\label{hermitic}\begin{split}
u(\xi,\eta,\lambda) A_0^{-1} u^\dag (\xi,\eta,\lambda^*) A_0 =\openone.
\end{split}\end{equation}
Let us now multiply (\ref{hermitic}) by $\lambda -\lambda_1$,  take the limit $\lambda \to \lambda_1$
and take into account eq. (\ref{eq:z3-uv}). This gives
\begin{equation}\label{eq:u5 ok}\begin{split}
 m_{k}=\frac{n_{4-k}^*}{\lambda_1^3 -\lambda_1^{*,3}}\sum_{ k=1}^{3}\kappa_{s-k} m_s m^*_{4-s}
\end{split}\end{equation}
where
\begin{equation}\label{eq:kapij0}\begin{split}
\kappa_0 = \lambda_1^{*,2}, \qquad  \kappa_1  = \lambda_1^{2}, \qquad  \kappa_2  = \lambda_1\lambda_1^{*}.
\end{split}\end{equation}
Thus, taking into account that $\lambda_1= \i\rho_1$, $\rho_1$ -- real and $m_k=m_k^*$, we obtain
\begin{equation}\label{eq:n-m}\begin{split}
n_1 &=\frac{2\lambda_1^3  m_3^* } {\lambda_1^2 m_3^*m_1 +  |\lambda_1|^2|m_2|^2 + \lambda_1^{2,*} m_1^*m_3  }
= \frac{ 2\i\rho_1 m_3}{2m_1m_3 - m_2^2} \\
n_2 &= \frac{ 2\lambda_1^3  m_2^* } {  \lambda_1^{2,*} m_3^*m_1 +  \lambda_1^2  |m_2|^2 + |\lambda_1|^2 m_1^*m_3  }
=  \frac{ 2\i\rho_1}{m_2} \\
n_3 &=  \frac{ 2\lambda_1^3  m_1^* } { |\lambda_1|^2 m_3^*m_1 +  \lambda_1^{2,*} |m_2|^2 + \lambda_1^2  m_1^*m_3  }
= \frac{ 2\i\rho_1 m_1}{m_2^2}.
\end{split}\end{equation}

In order to obtain  the vectors $|n\rangle$ and $\langle m^T|$ in terms of $\xi$ and $\eta$ we first impose the limit  $\lambda \to\lambda_1$
in equation (\ref{eq pt u}). We obtain that  the residue $A_1$ must satisfy
\begin{equation}\label{eq:A11'}\begin{split}
\i \frac{\partial A_1}{ \partial \xi } + 2 \i \phi_\xi H_0 A_1 + \lambda_1 [\mathcal{J}, A_1]=0 \\
\i \frac{\partial A_1}{ \partial \eta }  + \lambda_1^{-1} V_1 A_1 - \lambda_1^{-1} A_1 \mathcal{J}^2 =0.
\end{split}\end{equation}

Since $A_1 = |n \rangle \langle m^T|$ we find that (\ref{eq:A11'}) is satisfied if
\begin{equation}\label{eq:n1m1}\begin{aligned}
\i \frac{\partial |n\rangle }{ \partial \xi } + 2 \i \phi_\xi H_0 |n\rangle  + \lambda_1 \mathcal{J} |n\rangle &=0,
&\quad \i \frac{\partial \langle m^T| }{ \partial \xi } -\lambda_1 \langle m^T| \mathcal{J} &=0 \\
\i \frac{\partial |n\rangle }{ \partial \eta }  + \lambda_1^{-1} V_1|n\rangle &=0,
&\quad \i \frac{\partial \langle m^T| }{ \partial \eta } -\lambda_1^{-1} \langle m^T| \mathcal{J}^2 &=0
\end{aligned}\end{equation}
i.e.
\begin{equation}\label{eq:n1m2}\begin{split}
|n \rangle =\Psi(\xi,\eta, \i\rho_1) |n_{0}\rangle, \qquad \langle m^T| = \langle m^T_{0}|
(\Psi_0)^{-1} (\xi,\eta, \i\rho_1)
\end{split}\end{equation}
which means that  $|n (\xi, \eta)\rangle$ is an eigenfunction for the "dressed" Lax pair $L_1$, $L_2$, while $ \langle m^T(\xi, \eta)|$ is an eigenfunction for the "naked" Lax pair $L_{10}$, $L_{20}$.

From (\ref{Lax naked}), using direct calculation we obtain
\begin{equation}\label{eq:Psi00}\begin{split}
\Psi_0(\xi,\eta,\lambda) &= f_0  \e^{\i \lambda J\xi +\i \lambda^{-1} J^2 \eta} f_0^{-1} \\
f_0 = \frac{1}{\sqrt{3}} \left(\begin{array}{ccc} q & 1 & q^2 \\ 1 & 1 & 1 \\ q^2 & 1 & q
 \end{array}\right), \quad f_0^{-1}&= \frac{1}{\sqrt{3}} \left(\begin{array}{ccc} q^2 & 1 & q \\ 1 & 1 & 1 \\ q & 1 & q^2
 \end{array}\right),  \quad J = \diag (q^2,1,q)
\end{split}\end{equation}
which means that
\begin{equation}\label{eq:mxi}\begin{split}
\langle m^T| =\langle m^T_0| f_0 \e^{ \rho_1 J\xi -  \rho_1^{-1} J^2\eta} f_0^{-1}.
\end{split}\end{equation}
Using the notations
\begin{equation}\label{eq:mu00}\begin{split}
\langle m_{0}^T | \frac{1}{\sqrt{3}}f_0 = (\mu_{01}, \mu_{02}, \mu_{03})
\end{split}\end{equation}
we obtain the following explicit forms for  the components of vector $\langle m^T(\xi, \eta)|$
\begin{equation}\label{eq:mknou}\begin{split}
m_1(\xi,\eta) &= q^2\mu_{01} \e^{-\mathcal{X}_1} \e^{-\i \Omega_1} +\mu_{02} \e^{2 \mathcal{ X}_1}  + q\mu_{03} \e^{-\mathcal{X}_1} \e^{\i \Omega_1}  \\
m_2(\xi,\eta) &=  \mu_{01} \e^{-\mathcal{X}_1} \e^{-\i \Omega_1} +\mu_{02} \e^{2 \mathcal{ X}_1}  + \mu_{03} \e^{-\mathcal{X}_1} \e^{\i \Omega_1} \\
m_3(\xi,\eta) &=  q \mu_{01} \e^{-\mathcal{X}_1} \e^{-\i \Omega_1} +\mu_{02} \e^{2 \mathcal{ X}_1}  + q^2 \mu_{03} \e^{-\mathcal{X}_1} \e^{\i \Omega_1}
\end{split}\end{equation}
where
\begin{equation}\label{eq:X1}\begin{split}
\mathcal{X}_1 = \frac{1}{2} \left( \rho_1 \xi - \frac{\eta }{\rho_1} \right),  \qquad
\Omega_1 = \frac{\sqrt{3}}{2} \left( \rho_1 \xi + \frac{\eta }{\rho_1} \right).
\end{split}\end{equation}
For $\mu_{0,1} = \mu_{0,3}^* = |\mu_{01}|\e^{\i \alpha_0}$ and $\mu_{0,2} = \mu_{0,2}^*$ we can rewrite $m_i$ from (\ref{eq:mknou}) as the following real-valued functions
\begin{equation}\label{eq:new m}\begin{aligned}
m_1(\xi ,\eta) &=\mu_{02} \e^{2 \mathcal{X}_1} + 2| \mu_{01}|  \e^{-\mathcal{X}_1} \cos \left(   \Omega_1 - \alpha_{01} + \frac{2\pi}{3} \right) \\
m_2(\xi ,\eta) &=\mu_{02} \e^{2 \mathcal{X}_1} + 2|\mu_{01}| \e^{-\mathcal{X}_1} \cos \left(  \Omega_1 - \alpha_{01}  \right) \\
m_3 (\xi ,\eta) &=\mu_{02} \e^{2 \mathcal{X}_1} +2 |\mu_{01}| \e^{-\mathcal{X}_1} \cos \left( \Omega_1 - \alpha_{01} - \frac{2\pi}{3}  \right).
\end{aligned}\end{equation}
The components of the vector $|n\rangle $ in (\ref{eq:n-m}) become
 \begin{equation}\label{eq:n1}\begin{aligned}
 n_1 &= \frac{ 2\i\rho_1 m_3}{2m_1m_3-m_2^2}, &\quad  n_2 &= \frac{ 2\i\rho_1}{m_2}, &\quad
 n_3 &= \frac{ 2\i\rho_1 m_1}{m_2^2}.
 \end{aligned}\end{equation}

In order to obtain the solution of \c T2 equation we impose the limit $\lambda\to 0$ in (\ref{eq pt u}) with the result:

\begin{equation}\label{eq:phi0}\begin{split}
2\phi_\xi  \left(\begin{array}{ccc} 1 & 0 & 0 \\ 0 & 0 & 0 \\ 0 & 0 & -1 \end{array}\right)
=- \frac{\partial u}{ \partial \xi } u^{-1} (\xi,\eta,0)
\end{split}\end{equation}
where
\begin{equation}\label{eq:u0''}\begin{split}
u(\xi,\eta,0) &= \openone  -\frac{1}{3 \lambda_1} (A_1 +Q^{-1}A_1Q +Q^{-2}A_1Q^2 ) \\
&= \left( 1 - \frac{1}{\lambda_1} A_{1,jk}\right) \delta_{jk}
\end{split}\end{equation}
which means that
\begin{equation}\label{eq:phi1}\begin{split}
2\phi_\xi &=  -\frac{\partial u_{0;11}}{ \partial \xi }\frac{1}{u_{0;11}} = -\frac{\partial }{ \partial \xi } \ln u_{0;11}
\end{split}\end{equation}
or
\begin{equation}\label{sol 1}\begin{split}
2\phi  (\xi,\eta) &= -\ln \left| 1 - \frac{n_1 m_1 }{\lambda_1} \right| 
= \ln \left| \frac{2m_1m_3-m_2^2 }{m_2^2} \right| .
\end{split}\end{equation}

After introducing $m_i$ from (\ref{eq:new m}) we obtain the 1-soliton solution of the first type for $\lambda_1=\i \rho_1$
\begin{equation}\label{eq:sol TzII}\begin{split}
\phi_{1s}  (\xi,\eta) &=\frac{1}{2}\ln \left| \frac{|\mu_{01}|^2 \e^{-3\mathcal{X}_1} \left(  4\cos^2(\tilde{ \Omega}_1) -6\right) - 8|\mu_{01}|\mu_{02} \cos(\tilde{ \Omega}_1)
+\mu_{02}^2 \e^{3\mathcal{X}_1}}
{ 4|\mu_{01}|^2 \e^{-3\mathcal{X}_1} \cos^2(\tilde{ \Omega}_1) + 4|\mu_{01}|\mu_{02} \cos(\tilde{ \Omega}_1)  +\mu_{02}^2 \e^{3\mathcal{X}_1} }\right|
\end{split}\end{equation}
where $\tilde{ \Omega}_1=\Omega_1-\alpha_{01}$. 
We observe that this is not a traveling wave solution. Only if we take the limit $\mu_{02}\to 0$  we obtain a traveling wave solution of the form
\begin{equation}\label{eq:Mikhailov}\begin{split}
\phi  (\xi,\eta) &=\i \frac{\pi}{2}+\frac{1}{2}\ln \left[ \frac{3}{2}\tan^2\left(\frac{\sqrt{3}}{2}(\rho_1 \xi+\rho_1^{-1}\eta )-\alpha_{01}\right)+\frac{1}{2}\right].
\end{split}\end{equation}
The solution is singular and it blows up for $\frac{\sqrt{3}}{2}(\rho_1 \xi +\rho_1^{-1}\eta )-\alpha_{01}= (2k+1) \pi/2$, $k=0,\pm1,\dots$.
For $\alpha_{01} \rightarrow \alpha_{01}+\frac{\pi}{2}$ ($m_{10}, m_{20}, m_{30} \in \mathbb{C}$ and they are purely imaginary), the solution (\ref{eq:Mikhailov}) becomes
\begin{equation}\label{cotangenta}\begin{split}
\phi  (\xi,\eta) &=\i \frac{\pi}{2}+\frac{1}{2}\ln \left[\frac{3}{2}\cot^2\left(\frac{\sqrt{3}}{2}(\rho_1 \xi +\rho_1^{-1}\eta )-\alpha_{01}\right)+\frac{1}{2}\right].
\end{split}\end{equation}
The above solution is also singular and it blows up for $\frac{\sqrt{3}}{2}(\rho_1 \xi +\rho_1^{-1}\eta )+\alpha_0= k \pi$, $k=0,\pm1,\dots$.

\begin{remark}\label{rem:1}
It is easy to check, that the real parts of $\phi(\xi,\eta)$ in eqs. (\ref{eq:Mikhailov})  and  (\ref{cotangenta}) are in fact solutions to \c T4 equation.
\end{remark}

In order to get `quasi-regular' solutions of  \c T2 equation, we can apply the changes of variables $A_1^+$ with $a=\i$ or $A_2^+$ with
$b=\i$. This gives the following solutions  expressed in terms of hyperbolic functions
 \begin{equation}\label{eq:TzI tanh}\begin{split}
\phi  (\xi,\eta) &=\frac{1}{2}\ln \left[\frac{3}{2}\tanh^2\left(\frac{\sqrt{3}}{2}(\rho_1 \xi-\rho_1^{-1}\eta )-\alpha_{01}\right)-\frac{1}{2}\right]
\end{split}\end{equation}
and
 \begin{equation}\label{eq:TzI coth}\begin{split}
\phi  (\xi,\eta) &=\frac{1}{2}\ln \left[\frac{3}{2}\coth^2\left(\frac{\sqrt{3}}{2}(\rho_1 \xi -\rho_1^{-1}\eta )-\alpha_{01}\right)-\frac{1}{2}\right]
\end{split}\end{equation}
which are singular at the points for which
\[ \tanh \left(\frac{\sqrt{3}}{2}(\rho_1 \xi-\rho_1^{-1}\eta)-\alpha_{01}\right)= \pm \frac{1}{\sqrt{3}} \]
or
\[ \coth \left(\frac{\sqrt{3}}{2}(\rho_1 \xi-\rho_1^{-1}\eta )-\alpha_{01}\right)= \pm \frac{1}{\sqrt{3}} \]
respectively. These solutions have  also  been found by Mikhailov in \cite{mik}. As compared with the previous solutions, that have an infinite number of singularities,
these ones have singularities at only two points.  That is why we took the liberty to call them `quasi-regular'.

\subsection{The Singularity Properties of the Soliton Solutions}

Here we will discuss the types of singularities of the one-soliton solutions and how they are influenced by the changes of variables.
As we already mentioned, the singularities in the soliton solutions are not rare, see \cite{Fan,Mats}.

Let us first see how  the changes of variables affect the Lax representation (\ref{Lax dressed}) and, as a consequence,
how they affect the fundamental solution. We will be particularly  interested in the properties of the
`naked' Lax pair and its fundamental solution $\Psi_0(\xi,\eta,\lambda)$. This comes from the fact, that the soliton solution
is constructed as a rational function of the elements of $\Psi_0(\xi,\eta,\lambda)$.

Let us start with the change of variables $A_1^+$. Here the situation is simple; we readily get
\begin{equation}\label{eq:A1pLM}
\begin{split} L_1(\lambda)  \to \frac{1}{a} L_1(\lambda /a)  &\qquad 
 L_2(\lambda)  \to a L_2(a\lambda) \\
\Psi_0(\xi',\eta' ,\lambda')  &\to \Psi_0 \left( a\xi, \frac{\eta}{a} , \frac{ \lambda}{a}\right).
\end{split} \end{equation}
In other words this change of variables leaves invariant the compatibility of the Lax pair, so obviously it will map a
solution of \c T2 into a solution of \c T2. However now we have  to  keep in mind, that the change of variables must be
extended also to the spectral parameter $\lambda \to \lambda /a$ and, of course to the discrete eigenvalues  of $L_{1,2}$:
$\lambda_1 \to \lambda_1/a$ and therefore $\rho_1 \to  \rho_1 /a$.

In particular, from eq. (\ref{eq:X1}) we see, that
\begin{equation}\label{eq:X1'}\begin{aligned}
X_1'(\xi', \eta', \lambda_1') = \frac{1}{2} \left( \lambda_1' \xi' + \frac{\eta'}{\lambda_1'} \right) = X_1(\xi,\eta,\lambda_1) \\
\Omega_1'(\xi', \eta', \lambda_1') = \frac{1}{2} \left( \lambda_1' \xi' + \frac{\eta'}{\lambda_1'} \right) = \Omega_1(\xi,\eta,\lambda_1)
\end{aligned}\end{equation}
i.e. $X_1$ and $\Omega_1$ are invariant with respect to $A_1^+$  transformations provided
\begin{equation}\label{eq:la1}\begin{split}
 \lambda_1' = \frac{\lambda_1}{a}.
\end{split}\end{equation}

Now it is a bit more interesting to analyze the changes $A_2^+$.
\begin{equation}\label{eq:A2pLM}
\begin{split} L_1''(\lambda)  &\equiv  b \left( \i \frac{\partial }{ \partial \eta'' } + 2\i \phi_{\eta''} H_0 + \lambda  \mathcal{J}\right) \Psi(\xi '',\eta'',\lambda)=0 \\
L_2''(\lambda)  &\equiv b \left( \i \frac{\partial }{ \partial \xi'' } + \frac{ 1}{ \lambda b} V_1(\xi'',\eta'') \right)\Psi(\xi'',\eta'',\lambda)=0.
\end{split} \end{equation}
 Let us  apply a gauge transformation, i.e. change from $\Psi(\xi '',\eta'',\lambda)$ to
\begin{equation}\label{eq:A2LM3}\begin{split}
\tilde{ \Psi}(\xi'',\eta'',\lambda)  A_0 \e^{2\phi H_0} \Psi(\xi '',\eta'',\lambda'')
\end{split}\end{equation}
where $H_0$ and $A_0$ are defined in eqs. (\ref{eq:z2a-uv}) and (\ref{eq:z2b}) respectively. This gives us
\begin{equation}\label{eq:A2LM4}\begin{split}
L_1''(\lambda'') = L_2(\lambda), \qquad  L_2''(\lambda'') = L_1(\lambda),  \qquad \lambda'' = \frac{1}{b\lambda}.
\end{split}\end{equation}
So the $A_2^+$ change is equivalent to interchanging the Lax operators $L_1$ and $L_2$, which again preserves their compatibility condition.
Applied to $X_1$ and $\Phi_1$   these transformations lead to
\begin{equation}\label{eq:X1''}\begin{aligned}
\Psi_0''(\xi', \eta', \lambda_1'') = A_0\Psi_0(\xi,\eta,\lambda_1)A_0.
\end{aligned}\end{equation}

Of course, analyzing the fundamental solutions we have to pay attention also whether the parameters $a$ and $b$ are real or purely
imaginary. In addition we have to take into account, that $\lambda_1$ could be purely imaginary as above, but
for other cases it could also be real. It is precisely this choice
of the parameters $a$, $b$  and $\lambda_1$ that may change the singularity properties of the solutions.

\subsection{One Soliton Solutions of Second Type}

Our anzatz for the dressing factor is
\begin{equation}\label{eq:u06}\begin{split}
u(\xi,\eta,\lambda) = \openone &+ \frac{1}{3} \left(\frac{A_1 }{\lambda -\lambda_1} + \frac{Q^{-1}A_1Q }{\lambda q^2 -\lambda_1}  +
\frac{Q^{-2} A_1Q^2 }{\lambda q -\lambda_1} \right) \\
&- \frac{1}{3}\left( \frac{A_1^* }{\lambda +\lambda_1^*} + \frac{Q^{-1}A_1^*Q }{\lambda q^2 +\lambda_1^*}  +
\frac{Q^{-2} A_1^*Q^2 }{\lambda q +\lambda_1^*} \right)
\end{split}\end{equation}
which obviously satisfies the $\bbbz_3$-reduction and the first $\bbbz_2$-reduction.

In order to find how the components of the vector $|n\rangle$ are expressed in terms of the vector $|m^T\rangle$
we use the same procedure as in the $3$-poles case. First we rewrite the dressing factor in the following form.
\begin{equation}\label{eq:6p-01}\begin{split}
u(\xi,\eta,\lambda) =\openone + \frac{1}{\lambda^3 - \lambda_1^{3}}\mathcal{A}_1 (\xi,\eta,\lambda) -\frac{1}{\lambda^3 + \lambda_1^{3,*}}\mathcal{A}_1^* (\xi,\eta,-\lambda^*)
\end{split}\end{equation}
where
\begin{equation}\label{eq:6p-02}\begin{split}
\mathcal{A}_1 (\xi,\eta,\lambda) &=
\left(\begin{array}{ccc} \eta_0 n_1  m_1  & \eta_1 n_1  m_2
& \eta_2 n_1  m_3  \\ \eta_2 n_2  m_1  & \eta_0 n_2  m_2  & \eta_1 n_2  m_3  \\ \eta_1 n_3  m_1
& \eta_2 n_3  m_2  & \eta_0 n_3  m_3   \end{array}\right) \\
\mathcal{A}^*_1(\xi,\eta,-\lambda^*) &=
\left(\begin{array}{ccc} \rho_0 n_1^* m_1^* & \rho_1 n_1^* m_2^*
& \rho_2 n_1^* m_3^* \\ \rho_2 n_2^* m_1^* & \rho_0 n_2^* m_2^* & \rho_1 n_2^* m_3^* \\ \rho_1 n_3^* m_1^*
& \rho_2 n_3^* m_2^* & \rho_0 n_3^* m_3^*  \end{array}\right)
\end{split}\end{equation}
with
\begin{equation}\label{eq:6p-03}\begin{aligned}
\eta_0 & = \lambda_1^{2}, &\qquad \eta_1 & = \lambda \lambda_1, &\qquad \eta_2 & = \lambda^2 \\
\rho_0 & = \lambda_1^{*,2}, &\qquad \rho_1 & = -\lambda \lambda_1^*, &\qquad \rho_2 & = \lambda^2.
\end{aligned}\end{equation}
We insert the dressing factor $ u(\xi ,\eta ,\lambda ) $ into   The second $\bbbz_2$-reduction, we
multiply by $\lambda -\lambda_1$, take the limit $\lambda\to \lambda_1$ and obtain
\begin{equation}\label{eq:6 poles}\begin{split}
\langle m^T | A_0 &= \langle m^T | A_0\left[-\frac{1}{\lambda_1^3 - \lambda_1^{*,3}}  \mathcal{A}_{1}^\dag(\lambda_1^*)+\frac{1}{2\lambda_1^3 }\mathcal{A}_{1}^T(-\lambda_1)\right]
\end{split}\end{equation}
After direct calculation we obtain
\begin{equation}\label{eq:6p-05}\begin{aligned}
m_3 =  \zeta_1 K_1 n_1^* + c_1 P_1 n_1  \quad 
m_2 =  \zeta_1 K_2 n_2^* + c_1 P_2 n_2 \quad 
m_1 =  \zeta_1 K_3 n_3^* + c_1 P_3 n_3
\end{aligned}\end{equation}
where
\begin{equation}\label{eq:6p-06}\begin{aligned}
K_1 & = \lambda_1^{*,2} m_3 m_1^* + \lambda_1 \lambda_1^* m_2m_2^* +\lambda_1^2 m_1m_3^*, &\; P_1 &= 2m_1m_3 -m_2^2 \\
K_2 & = \lambda_1^{2} m_3 m_1^* +  \lambda_1^{*,2} m_2m_2^* +\lambda_1\lambda_1^* m_1m_3^*, &\; P_2 &= m_2^2 \\
K_3 & = \lambda_1\lambda_1^{*} m_3 m_1^* + \lambda_1^2 m_2m_2^* +\lambda_1^{*,2} m_1m_3^*, &\; P_3 &= m_2^2 \\
\zeta_1 & =- \frac{1}{\lambda_1^3 - \lambda_1^{*,3}}, &\; c_1 &= \frac{1}{2\lambda_1}.
\end{aligned}\end{equation}
We rewrite the result in a matrix form
\begin{equation}\label{eq:6p-07}\begin{split}
|\mu\rangle=\left(\begin{array}{c} m_3 \\ m_2 \\ m_1 \\ \hline m_3^* \\ m_2^* \\ m_1^*  \end{array}\right), \qquad
|\nu\rangle=\left(\begin{array}{c} n_1 \\ n_2 \\ n_3 \\ \hline n_1^* \\ n_2^* \\ n_3^*  \end{array}\right), \qquad
|\mu\rangle=\mathcal{M} |\nu\rangle
\end{split}\end{equation}
where
\begin{equation}\label{eq:miu and niu}\begin{split}
\mathcal{M}=\left(\begin{array}{ccc|ccc} c_1 P_1 & 0 & 0 & \zeta_1 K_1 & 0 & 0 \\ 0& c_1 P_2 &  0 &  0 & \zeta_1 K_2 & 0  \\
0 & 0 & c_1 P_3 & 0 & 0 & \zeta_1 K_3  \\ \hline \zeta_1 K_1^* & 0 & 0 & c_1 P_1^* & 0 & 0 \\ 0 & \zeta_1 K_2^* & 0 & 0 & c_1 P_2^* &  0 \\
0 & 0 & \zeta_1 K_3^* & 0 & 0 & c_1 P_3^*   \end{array}\right)
\end{split}\end{equation}
The result is
\begin{equation}\label{eq:6p-07'}\begin{split}
|\nu\rangle & =\mathcal{M}^{-1} |\nu\rangle \\
\mathcal{M}^{-1}  &=
\left(\begin{array}{ccc|ccc} -c_1^* \tilde{P}_1^* & 0 & 0 & \zeta_1 \tilde{K}_1 & 0 & 0 \\ 0& -c_1^* \tilde{P}_2^* &  0 &  0 & \zeta_1 \tilde{K}_2 & 0  \\
0 & 0 & -c_1^* \tilde{P}_3^* & 0 & 0 & \zeta_1 \tilde{K}_3 \\ \hline \zeta_1^* \tilde{K}_1^* & 0 & 0 & -c_1 \tilde{P}_1 & 0 & 0
\\ 0 & \zeta_1^* \tilde{K}_2^* & 0 & 0 & -c_1 \tilde{P}_2 &  0 \\ 0 & 0 & \zeta_1^* \tilde{K}_3^* & 0 & 0 & -c_1 \tilde{P}_3   \end{array}\right)
\end{split}\end{equation}
where
\[ \tilde{P}_s^* = \frac{P_s^*}{d_s},  \qquad \tilde{P}_s = \frac{P_s}{d_s}, \qquad  \tilde{K}_s = \frac{K_s}{d_s},
\qquad \tilde{K}_s^* = \frac{K_s^*}{d_1} \]
\begin{equation}\label{eq:mn10}\begin{aligned}
d_1 = \zeta_1 \zeta_1^* K_1K_1^* &- c_1c_1^* P_1P_1^*  \qquad 
 d_2  = \zeta_1 \zeta_1^* K_2K_2^* - c_1c_1^* P_2P_2^* \\ 
d_3 &= \zeta_1 \zeta_1^* K_3K_3^* - c_1c_1^* P_3P_3^*.
\end{aligned}\end{equation}
From the above equations we obtain $|n\rangle$ in terms of $\langle m^T|$
\begin{equation}\label{eq:6p-05'}\begin{aligned}
n_1 =  \frac{1}{d_1}(-c_1^* P_1^* m_3& +\zeta_1 K_1 m_3^*)  \qquad
n_2 =  \frac{1}{d_2}(-c_1^* P_2^* m_2+\zeta_1 K_2 m_2^*) \\
n_3 &=  \frac{1}{d_3}(-c_1^* P_3^* m_1+\zeta_1 K_3 m_1^*).
\end{aligned}\end{equation}

In this case we choose a general form for the poles: $\lambda_1 =\rho_1 \e^{\i \beta_1}$; without restrictions we can choose $0 <\beta_1 < \frac{\pi}{6}$
and determine the expressions of $\langle m^T|$ as
\begin{equation}\label{eq:m10'}\begin{split}
m_1&=q^2\mu_{01}\e^{\i \mathcal{X}_1-\mathcal{Y}_1}+\mu_{02}\e^{\i \mathcal{X}_2-\mathcal{Y}_2}+q\mu_{03}\e^{\i \mathcal{X}_3- \mathcal{Y}_3} \\
m_2 &= \mu_{01} e^{\i \mathcal{X}_1 - \mathcal{Y}_1} + \mu_{02} e^{\i \mathcal{X}_2 -\mathcal{Y}_2} + \mu_{03} \e^{\i \mathcal{X}_3 - \mathcal{Y}_3}  \\
m_3 &= q\mu_{01} \e^{\i \mathcal{X}_1 - \mathcal{Y}_1} + \mu_{02} \e^{\i \mathcal{X}_2 - \mathcal{Y}_2} + q^2\mu_{03} \e^{\i \mathcal{X}_3 - \mathcal{Y}_3}
\end{split}\end{equation}
where
\begin{equation}\label{eq:Ee3}\begin{aligned}
\mathcal{X}_1 &=- \left( \xi \rho_1 + \frac{\eta}{\rho_1} \right) \cos \left(\beta_1 - \frac{2\pi}{3}\right), &\; \mathcal{Y}_1 &=-
 \left( \xi \rho_1 - \frac{\eta}{\rho_1} \right) \sin \left(\beta_1 - \frac{2\pi}{3}\right)\\
\mathcal{X}_2 &= -\left(\xi \rho_1 + \frac{\eta}{\rho_1} \right) \cos \left(\beta_1 \right), &\;
\mathcal{Y}_2 &= -\left( \xi \rho_1 - \frac{\eta}{\rho_1} \right) \sin \left(\beta_1 \right)\\
\mathcal{X}_3 &= -\left( \xi \rho_1 + \frac{\eta}{\rho_1} \right) \cos \left(\beta_1 + \frac{2\pi}{3}\right), &\;
\mathcal{Y}_3 &=- \left( \xi \rho_1 - \frac{\eta}{\rho_1} \right) \sin \left(\beta_1 + \frac{2\pi}{3}\right).
\end{aligned}\end{equation}
We determine the 1-soliton solution for the second kind of solitons using exactly the same technique
\begin{equation}
 \Phi=-\frac{1}{2}\ln \left|1-\frac{1}{\lambda _{1}}n_{1}m_{1}-\frac{1}{\lambda^*_{1}} n^*_{1}m ^*_{1}\right|.
\end{equation}

\section{The Generic $N$-soliton Solution for \c T2 Equation}

Let us consider the dressing factor of the following form
\begin{equation}\label{eq:6NuNp''m}\begin{split}
 u(\xi ,\eta ,\lambda ) &=\openone + \frac{1}{3}\sum_{s=0}^{2} \left(
 \sum_{l=1}^{N_1} \frac{ Q^{-s}A_{l} Q^s }{\lambda q^s-\lambda_l}+\sum_{r=N_1+1}^{N}
\left( \frac{ Q^{-s} A_{r} Q^s}{q^s\lambda -\lambda_r }-  \frac{Q^{-s} A^*_{r}Q^s}{\lambda q^s+\lambda_r^* } \right)\right)
\end{split}\end{equation}
with  $3N_1 +6N_2$ complex poles
from which $N_1$ are purely imaginary, satisfying the following condition: $\lambda_p=-\lambda_p^*$.

Then we write down the residues $A_k(\xi,\eta)$ as degenerate $3 \times 3$ matrices of the form
\begin{equation} \begin{split}
 A_k(\xi,\eta) = |n_k(\xi,\eta)\rangle \langle m_k^T(\xi,\eta)|, \qquad
(A_k)_{ij}(\xi,\eta) = n_{ki}(\xi,\eta)m_{kj}(\xi,\eta).
\end{split}\end{equation}
From  the second $\bbbz_2$-reduction, $A_0^{-1} u^\dag (\xi, \eta,\lambda^* )A_0 = u^{-1} (\xi,\eta, \lambda )$,
after taking the limit $\lambda \to\lambda_k$, we obtain  algebraic equation for $|n_k\rangle$ in terms of $\langle m_k^T|$
\begin{equation}\begin{split}
|\nu\rangle = \mathcal{M}^{-1}|\mu\rangle .
\end{split}\end{equation}
Below, for simplicity, we write down the matrix $\mathcal{M}$ for $N_1=N_2=1$
\begin{equation}\begin{split}
 |\nu\rangle=\left(\begin{array}{c} |n_1\rangle \\ \hline  |n_2\rangle\\  |n^*_2\rangle \end{array}\right), \qquad
|\mu \rangle =\left(\begin{array}{c} A_0|m_1\rangle \\ \hline  A_0|m_2\rangle\\  A_0|m^*_2\rangle \end{array}\right),
\qquad  \mathcal{M} & =\left(\begin{array}{c|cc} A &  B &  B^*\\ \hline B  &  D  & E  \\- B^* & -E^*  & D^*  \end{array}\right)
\end{split}\end{equation}
\begin{equation}\label{eq:6N2s3mm}\begin{aligned}
A  &=\frac {1}{2\lambda_1^3 } \diag (Q^{(1)} , Q^{(2)} , Q^{(3)} ), &\quad
B  &=\frac {1}{\lambda_1^3+ \lambda_2^3} \diag (P^{(1)} , P^{(2)} , P^{(3)} ) \\
D  &=\frac {1}{2\lambda_2^3 } \diag (T^{(1)}, T^{(2)}, T^{(3)}), &\quad
E  &=\frac {1}{\lambda_2^{*,3} - \lambda_2^3} \diag (K^{(1)},K^{(2)}, K^{(3)}) \\
Q^{(j)}  & = \langle m_1^T| \Lambda_{11}^{(j)}(\lambda_1, \lambda_1) |m_1 \rangle , &\; P^{(j)}  & = \langle m_1^T| \Lambda_{12}^{(j)}(\lambda_1, \lambda_2) |m_2 \rangle \\
T^{(j)}  & = \langle m_2^T| \Lambda_{22}^{(j)}(\lambda_2, \lambda_2) |m_2 \rangle , &\; K^{(j)}   &= \langle m_2^T| \Lambda_{22}^{(j)}(\lambda_2, -\lambda^*_2) |m_2^* \rangle
\end{aligned}\end{equation}
with
\begin{equation}\label{eq:QKP}\begin{aligned}
\Lambda^{(j)}_{lp} & =-\lambda_l \lambda_p E_{3-j, 1+j}+\lambda_l^2 E_{2-j,2+j}+\lambda_p^2 E_{1-j,3+j},  \quad j=1,2,3.
\end{aligned}\end{equation}
In order to obtain the 2-soliton solution of the Tzitzeica equation  we take the limit
$\lambda\to 0$ in the equations satisfied by the dressing factor $u(\xi,\eta, \lambda)$ and integrate to get
\begin{equation}\label{eq:6N mi}
\phi_{\rm Ns}(\xi,\eta) = -\frac{1}{2}\ln \left| 1 - \frac{n_{1,1} m_{1,1} }{\lambda_1}-
\frac{n_{2,1} m_{2,1} }{\lambda_2}-  \frac{n^*_{2,1} m^*_{2,1} }{\lambda^*_2}\right|.
\end{equation}
The above formulae can be easily generalized for any $N_1$ and $N_2$.

\section{Hirota Method for Building 1-soliton Solution of \c T2 Equation}

There are many methods for deriving the soliton solutions; we have demonstrated two of the most used: the dressing method and the
Hirota method \cite{hir2002,cori5,noi}. Both methods give the same results both for the kinks and for the breathers.

We build the Hirota bilinear form of \c T2 eq. using the substitution
\begin{equation}
\phi  (\xi,\eta)=\frac{1}{2}\ln\frac{g(\xi,\eta)}{f(\xi,\eta)}.
\end{equation}
 Introducing it into the second equation in  (\ref{eq:Ts12}) and decoupling in
the bilinear dispersion relation and the soliton-phase constraint, we obtain the following system
\begin{equation}\label{eq:fg0}\begin{split}
\frac{\partial^2 g  }{\partial \xi \partial \eta} g - \frac{\partial g}{ \partial \xi }\frac{\partial g}{ \partial \eta } -f^2+g^2=0 \\
\frac{\partial^2 f  }{\partial \xi \partial \eta} g - \frac{\partial f}{ \partial \xi }\frac{\partial f}{ \partial \eta } -fg+ f^2=0.
\end{split}\end{equation}
We impose that:
\begin{equation}\label{ansatz}\begin{split}
g(\xi, \eta)=1+a z(\xi,\eta)+ b z^2(\xi, \eta)\\
f(\xi, \eta)=1+A z(\xi,\eta)+ B z^2(\xi, \eta)
\end{split}\end{equation}
where $z(\xi,\eta)=e^{k \xi-\omega \eta}$, $ k$ - the wave number, $\omega$ - the angular frequency.

Using a software for analytical computation like MATHEMATICA, we obtain that
\begin{equation}\label{sol Hirota}\begin{split}
g(\xi, \eta)=1-2A \e^{k \xi-\frac{3}{k}\eta}+\frac{A^2}{4} \e^{k \xi-\frac{3}{k}\eta}\\
f(\xi, \eta)=1+A \e^{k \xi-\frac{3}{k}\eta}+\frac{A^2}{4} \e^{k \xi-\frac{3}{k}\eta}
\end{split}\end{equation}
where the dispersion relation is $\omega=\frac{3}{k}$.

Using the above results our 1-soliton solution for \c T2 acquires the following form
\begin{equation}\label{Hirota}\begin{split}
\phi  (\xi,\eta) &=\frac{1}{2}\ln \left[ \frac{3}{2}\tanh^2\left(\frac{1}{2}(k\xi-\frac{3}{k}\eta)\right)-\frac{1}{2}\right] .
\end{split}\end{equation}

This solution coincide with the one obtained by Mikhailov in \cite{mik} for $k= \sqrt{3}\rho_1$. In this very direct manner,
Hirota method gives immediately the 1-soliton solution of the first type, which we have obtained also in (\ref{eq:TzI tanh})
through  the dressing method, as a particular case of a more general form (\ref{eq:sol TzII}).

One can also use the standard Hirota technique to derive  $N$-soliton solution of first type each parametrized
with real eigenvalue $\rho_k$ and a vector
$(\mu_{k,1}, \mu_{k,2}, \mu_{k,3})$  with $\mu_{k,2}=0 $. We believe,
that using Hirota method one can derive also more complicated cases of one- and $N$-soliton solutions. To this end, however
one needs a more complicated ansatz  for the functions $f(\xi, \eta)$ and $g(\xi, \eta)$ which would solve equation (\ref{eq:fg0}) but could not be reduced
to functions of $z(\xi,\eta)$ only.

Of course, the equation (\ref{eq:fg0}) can be solved in a more general case, but the only one solution we were able to obtain by now, using the well known ansatz (\ref{ansatz}), was (\ref{sol Hirota}), which corresponds to the soliton solutions of first type. To find $g(\xi, \eta)$ and $f(\xi, \eta)$ corresponding to the second type of soliton solutions is still an open problem for us and it will be tackled in a next paper.
A possible approach could be to start from the second type solitons given by the dressing factor method and, on
this basis, to guess the ansatz which should be imposed to obtain $g(\xi, \eta)$ and $f(\xi, \eta)$ verifying (\ref{eq:fg0}).

\section{The Spectral Properties of the Dressed Lax  Operator}

Here we shall demonstrate that each dressing adds to the discrete spectrum of $L$ sets of discrete eigenvalues.

In our previous paper we showed that the Lax operator has a set of 6 fundamental analytic solutions. We will denote them
by $\chi_\nu (\xi,\eta ,\lambda)$ where $\nu=0,\dots, 5$ denotes the number of the sector $\Omega_\nu \equiv \frac{(2\nu +1)\pi }{6}
\leq \arg \lambda \leq \frac{ (2\nu +3)\pi }{6} $, i.e. those are the sectors closed by the rays $(l_\nu , l_{\nu +1} )$.
The dressing factor for solitons of first type  (\ref{eq:u0})  obviously has simple poles located at $|\lambda_1| \e^{2\pi \i k/3}$, $k=0,1,2$.
The inverse of this dressing factor has also simple poles located at $|\lambda_1| e^{\pi \i (2 k+1)/3}$, $k=0,1,2$.
Each dressing factor for soliton of second type  (\ref{eq:u06})   has 6 simple poles located at $|\lambda_2| \e^{\i \beta_1 +2\pi \i k/3}$
and $|\lambda_2| \e^{-\i \beta_1 +\pi \i(2 k+1)/3}$, $k=0,1,2$. The inverse of this dressing factor has also 6 simple poles located at
 $|\lambda_2| \e^{\i \beta_1 +\pi \i(2 k+1)/3}$ and $|\lambda_2| \e^{-\i \beta_1 +2\pi \i  k/3}$, $k=0,1,2$.

\begin{figure}[t]
\centering
\setlength{\unitlength}{0.1800pt}
\ifx\plotpoint\undefined\newsavebox{\plotpoint}\fi
\sbox{\plotpoint}{\rule[-0.175pt]{0.350pt}{0.350pt}}%
\special{em:linewidth 0.5pt}%
\begin{picture}(1440,1440)(0,0)

\put(1160.,1273){\Large $\lambda$}
\put(720,220){\line(0,1){1000}} %\put(1260,720){\LARGE $l_1$}

\put(720,720){\line(2,1){500}} %\put(1260,720){\LARGE $l_1$}

\put(720,720){\line(-2,1){500}} %\put(1260,720){\LARGE $l_1$}

\put(720,720){\line(2,-1){500}} %\put(1260,720){\LARGE $l_1$}

\put(720,720){\line(-2,-1){500}} %\put(1260,720){\LARGE $l_1$}

\put(830,936.5){${}_\times$} \put(955,720){${}_\otimes$}

\put(455,720){${}_\times$} \put(580,936.5){${}_\otimes$}

\put(830,503.5){${}_\times$} \put(580,503.5){${}_\otimes$}

\dottedline{10}(220,720)(1220,720)

\dottedline{10}(470,287)(970,1153)

\dottedline{10}(970,287)(470,1153)
%%%%%%%%%%%%%%%%%%%%%% PLUS \pi/10
\put(767.4,1013.4){${}_+$} \put(990.3,812.7){${}_\oplus$}

\put(419.7,627.3){${}_+$} \put(482.1,920.7){${}_\oplus$}

\put(927.9,519.3){${}_+$} \put(642.6,426.6){${}_\oplus$}

%%%%%%%%%%%%%%%%%%%%%% MINUS \pi/10

\put(927.9,920.7){${}_+$} \put(990.3,627.3){${}_\oplus$}

\put(419.7,812.7){${}_+$} \put(642.6,1013.4){${}_\oplus$}

\put(767.4,426.6){${}_+$} \put(482.1,519.3){${}_\oplus$}
\put(720,1260){\large $l_1$}
\put(1173,980.){\large $l_0$}
\put(1173,495.){\large $l_5$}
 \put(720,180.){\large $l_4$}
\put(250,450.){\large $l_3$}
\put(250,950.){\large $l_2$}

\put(970,1153){\large $b_0$}
\put(1220,720){\large $b_5$}
\put(970,287){\large $b_4$}
\put(470,287){\large $b_3$}
\put(220,720){\large $b_2$}
\put(470,1153){\large $b_1$}
\end{picture}
\caption{The contour of the RHP with $\bbbz_3$-symmetry fills up the rays $l_0$, \dots , $l_ 5$.
The symbols $\times$ and $\otimes$ (resp.  $+$ and $\oplus$)  denote the locations of the
discrete eigenvalues corresponding to a soliton of first (resp.  second) type.
\label{fig:1}}

\end{figure}
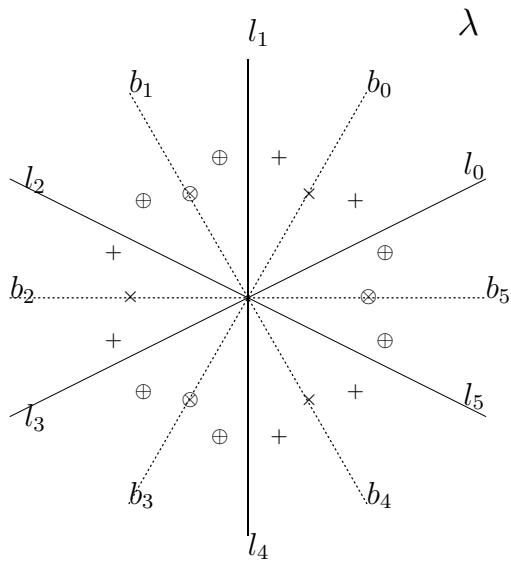

The FAS can be used to construct the kernel of the resolvent of the Lax operator $L$.
In this section by $\chi^\nu (\xi,\lambda)$ we will denote
\begin{equation}\label{eq:1}\begin{split}
\chi^\nu (\xi,\lambda) = u(\xi,\lambda) \chi_0^\nu (\xi,\lambda) u_-^{-1}(\lambda), \qquad u_-(\lambda) =\lim_{\xi\to -\infty}
u(\xi,\eta,\lambda)
\end{split}\end{equation}
where $\chi_0^\nu (\xi,\lambda)$ is a regular FAS and $u(\xi,\lambda)$ is a dressing factor of
general form (\ref{eq:6NuNp''m}).
Let us introduce
\begin{equation}\label{eq:R-nu}\begin{aligned}
R^\nu (\xi, \xi',\lambda) &= \frac{1}{\i} \chi^\nu (\xi,\lambda) \Theta_\nu (\xi -\xi') \hat{\chi}^\nu (\xi',\lambda)
\end{aligned}\end{equation}
\begin{equation}\label{eq:Theta}\begin{split}
\Theta_\nu (\xi - \xi') = \diag \left( \eta_\nu^{(1)} \theta( \eta_\nu^{(1)} (\xi -\xi')),
\eta_\nu^{(2)} \theta( \eta_\nu^{(2)} (\xi -\xi')), \eta_\nu^{(3)} \theta( \eta_\nu^{(3)} (\xi -\xi')) \right)
\end{split}\end{equation}
where  $\theta(\xi -\xi')$ is the step-function and $\eta_\nu^{(k)}=\pm 1$, see the table \ref{tab:2}.
\begin{table}
\centering
\begin{tabular}{|c|c|c|c|c|c|c|}
  \hline
  % after \\: \hline or \cline{col1-col2} \cline{col3-col4} ...
  $ $ & $\Upsilon_0$  &  $\Upsilon_1$ &  $\Upsilon_2$ &  $\Upsilon_3$ &  $\Upsilon_4$ &  $\Upsilon_5$ \\ \hline
  $\eta_{\nu}^{(1)} $ & $-$ & $-$ & $-$ & $+$ & $+$ & $+$ \\
  $\eta_{\nu}^{(2)}$ &  $+$ & $+$ & $-$ & $-$ & $-$ & $+$ \\
  $\eta_{\nu}^{(3)}$ & $-$ & $+$ & $+$ & $+$ & $-$ & $-$ \\%%
  \hline
\end{tabular}
\caption{The set of signs $\eta_{\nu}^{(k)}$ for each of the sectors $\Upsilon_\nu$ (\ref{eq:Upsi}). \label{tab:2}}
\end{table}

Then the following theorem holds true \cite{BJGA}:
\begin{theorem}\label{thm:1}
Let $Q(\xi) $ be a Schwartz-type function  and let $ \lambda _j^\pm $
be the simple zeroes of the dressing factor $u(\xi, \lambda ) $ (\ref{eq:6NuNp''m}). Then

\begin{enumerate}

\item The functions $R^\nu (\xi, \xi',\lambda)$ are analytic for $\lambda\in\Upsilon_\nu$ where
\begin{equation}\label{eq:Upsi}\begin{split}
b_\nu \colon \arg \lambda = \frac{\pi (\nu+1)}{3} , \qquad \Upsilon_\nu \colon
\frac{\pi (\nu+1)}{3} \leq \arg \lambda \leq \frac{\pi (\nu+2)}{3}
\end{split}\end{equation}
 having pole singularities at $\pm \lambda _j^\pm $;

\item $R^\nu (\xi, \xi',\lambda ) $ is a kernel of a bounded integral operator
for $\lambda \in \Upsilon_\nu $;

\item $R^\nu (\xi, \xi',\lambda ) $ is uniformly bounded function for $\lambda
\in b_\nu $ and provides a kernel of an unbounded integral operator;

\item $R^\nu (\xi, \xi',\lambda ) $ satisfy the equation
\begin{equation}\label{eq:R3.1}
L(\lambda ) R^\nu (\xi, \xi',\lambda )=\openone \delta (\xi- \xi').
\end{equation}
\end{enumerate}
\end{theorem}

\begin{remark}\label{rem:2}
The dressing factor $u(\xi,\lambda)$ has $3N_1+6N_2$ simple poles located at
$\lambda_l q^p$, $\lambda_r q^p$ and $\lambda_r^* q^p$ where $l=1,\dots, N_1$, $r=1,\dots , N_2$ and $p=0,1,2$.
Its inverse $u^{-1}(\xi,\lambda)$ has also $3N_1+6N_2$ poles located $-\lambda_l q^p$, $-\lambda_r q^p$ and $-\lambda_r^* q^p$.
In what follows for brevity we will denote them by $\lambda_j $, $-\lambda_j$ for $j=1,\dots , 3N_1+6N_2$.
\end{remark}

It remains to show that the poles of $R^\nu (\xi, \xi',\lambda)$ coincide with
the poles of the dressing factors $u(\xi,\lambda)$ and its inverse $u^{-1}(\xi,\lambda)$.

The proof follows immediately from the definition of $R^\nu (\xi, \xi',\lambda)$ and from eq. (\ref{eq:1}), taking into account that
the limiting value $u_-(\lambda)$ commutes with the corresponding matrix $\Theta_\nu (\xi-\xi')$.

Thus we have established that dressing by the factor $u(\xi,\lambda)$, we in fact add to the discrete spectrum
of the Lax operator $6N_1+12N_2$ discrete eigenvalues; for $N_1=N_2=1$ they are shown on Figure \ref{fig:1}.

\section{Conclusions}

Shortly before finishing this paper we became aware of the fact, that appropriate combination of changes of variables,
considered in Section 2 can take each member of Tzitzeica family (\ref{eq:TzF}) into one of its 4 versions that we introduced.
Let us demonstrate how this can be done of the equation
\begin{equation}\label{eq:TsF2}\begin{split}
\frac{\partial ^2 \phi}{ \partial \xi \partial \eta } = - c_1^2 \e^{2\phi} + c_2^2 \e^{-4\phi}
\end{split}\end{equation}
where $c_1$ and $c_2$ are real positive constants. Now we shall use somewhat more general change of variables. First we apply
the transformation (\ref{eq:phi'}) with $s_0=0$ and $\phi' =\phi +\ln (c_1/c_2)$. Then we change  $\xi \to \xi' /k$,
$\eta \to \eta'/k$ where $k$ is also real positive constant taken to be $k= \sqrt[3]{c_1^2 c_2}$. Easy calculation shows that as a
result eq. (\ref{eq:TsF2}) goes into \c T2 for $\phi'(\xi,\eta)$.
Using in addition Table \ref{tab:1} we can transform each member of Tzitzeica family into \c T2 and then solve it using the results above.

 Let us consider the soliton solutions Tzitzeica eq. in a small neighborhood  around the singularities, where  $\phi_{\rm as}(\xi,\eta)$ tends to $\infty$.
Then the second term in the \c T2 equation can be neglected and the asymptotically we get
\[ 2\frac{\partial \phi_{\rm as}}{ \partial \xi  \partial \eta} = \e^{2\phi_{\rm as}}. \]
Similarly if in a small neighborhood  around the singularity   $\phi_{\rm as}'(\xi,\eta)$ tends to $-\infty$ we have
\[ 2\frac{\partial \phi_{\rm as}'}{ \partial \xi  \partial \eta} = - \e^{-4\phi_{\rm as}'}. \]
In both cases we find equations, equivalent to the Liouville equation.
Thus the asymptotical behavior of the solutions of Tzitzeica equation around the singularities
must be the same as the singularities of Liouville equation \cite{Pogr2}.

\section*{Acknowledgements}
One of us (VSG) is grateful to professor A. V. Mikhailov and professor A. K. Pogrebkov for useful discussions.
This work has been supported in part by a joint project between the Bulgarian academy of sciences and the Romanian academy
of sciences. One of the authors (CNB)  acknowledges the support of the strategic grant POSDRU/159/1.5/S/133255,
Project ID 133255 (2014), co-financed by the European Social Fund within the Sectorial Operational Program Human Resources
Development 2007-2013, and also the support of the project IDEI, PN-II-ID-PCE-2011-3-0083 (MECTS).

%Insert here your data,  e.g.\\
Corina N. Babalic \\
Dept. of Physics \\ University of Craiova \\ St. Alexandru Ioan Cuza 13\\
200585 Craiova,  Romania\\ and \\
%Dept. of Theoretical Physics \\
NIPNE, Str. Reactorului 30\\
Magurele, Bucharest, Romania \\
{\it E-mail address}: {\tt b$\underline{\;\;}$coryna@yahoo.com}\\[0.3cm]
 Radu Constantinescu \\
Dept. of Physics \\ University of Craiova \\ St. Alexandru Ioan Cuza 13\\
200585 Craiova,  Romania\\
{\it E-mail address}: {\tt rconsta@yahoo.com}\\[0.3cm]
Vladimir S. Gerdjikov \\
Institute for Nuclear Research and Nuclear Energy \\
Bulgarian Academy of Sciences \\
1784 Sofia, Bulgaria \\
{\it E-mail address}: {\tt gerjikov@inrne.bas.bg}

\label{last}
\end{document}

%% file: somedef.tex
\def\diag{
{\rm diag}}

\def\tan{
{\rm tan}}

\newcommand{\e}{\mathrm{e}}
\renewcommand{\i}{\mathrm{i}}

\theoremstyle{plain}
\newtheorem{theorem}{Theorem}

\newtheorem{remark}[theorem]{Remark}